\RequirePackage{ifpdf}
\documentclass{PoS}

\title{Early Phase Detection and Coverage of Extragalactic and Galactic
Black Hole X-ray Transients with the SKA}

\ShortTitle{Early Phase Coverage of Black Hole X-ray Transients on All Scales}

\author{\speaker{Wenfei Yu}\thanks{We acknowledge support by the National Natural Science Foundation of China under grant No. 11333005.
}, Hui Zhang, Zhen Yan and Wenda Zhang\\
	Shanghai Astronomical Observatory\\
	80 Nandan Road, Shanghai 200030, China\\
        E-mail: \email{wenfei@shao.ac.cn}}

\abstract{ SKA's large field of view and high sensitivity at low
frequencies will provide almost a complete coverage of the very early
rising phase of extragalactic and Galactic transients which undergo a
flare or outburst due to an abrupt accretion onto either supermassive (such
as tidal disruption events, TDEs) or stellar mass black hole transients  (such as
black hole LMXB) , when their broadband emission is supposed to
be jet-dominated at low luminosities, allowing SKA to be the first facility
to make source discoveries and to send out alerts for follow-up ground or
space observations as compared with the sensitivity of future X-ray
wide-field-view monitoring. On the other hand, due to extremely large
rate-of-change in the mass accretion rate during the rising phase of TDE
flares or transient outbursts, SKA will be able to cover an extremely large
range of the mass accretion rate as well as its rate-of-change not accessible with observations in persistent black hole
systems, which will shape our understanding of disk-jet coupling in
accreting black holes in the non-stationary accretion regimes.  }

\FullConference{
Advancing Astrophysics with the Square Kilometre Array\\
June 8-13, 2014\\
Giardini Naxos, Italy}

\newcommand{\skipthis}[1]{}
\newcommand\nar{New Astronomy Reviews}
\newcommand\apj{ApJ}
\newcommand\mnras{MNRAS}
\newcommand\apss{APSS}
\newcommand\nat{Nature}

\begin{document}

\section{Scientific background}

Wide field-of-view (FOV) astronomical monitors and survey telescopes are
very important to the time-domain astronomy, a rapidly developing field
with the advances of space and ground telescopes equiped with either fast
response or wide FOV monitoring or survey capabilities across a broad range
of the electromagnetic wavelength. Diverse time domain phenomena, such as
gamma-ray bursts, supernova explosions, tidal disruption flares, X-ray
nova, giant flares from soft gamma repeater, transient pulsars or fast
radio bursts, and so on, can only be efficiently detected with sensitive
monitoring or wide FOV surveying telescopes. Wide FOV monitoring and
surveying telescopes will bring a great impact on future astronomy.  

All-sky or wide FOV X-ray monitoring has led such time domain studies on
high energy transients for several decades. The success can be dated back
as early as the discovery of gamma-ray bursts which was about 50 years ago.
Current X-ray monitors remain at the frontier of time-domain astronomy on
high energy transients. For instance, new scientific discoveries continue
to occur from monitoring or survey observations made with space
observatories such as Swift, MAXI, and INTEGRAL. However, newly built or
currently planned multi-wavelength time-domain ground facilities, such as
the Palomer Transient Factory (PTF), the Pan-SSTARs, the LSST, and the
LOFAR etc., will bring time domain astrophysics to a golden era, allowing
almost simultaneous broadband coverage of those transient events,
including potential transient events associated with strong gravitational
waves. Here we emphasize that wide FOV monitoring observations with the SKA
have significant advantages in observing Galactic and extragalactic black
hole transients in the early rising phase, and have a great potential to take over
future space X-ray monitors in monitoring Galactic and extragalactic black
hole transients based on what we have learnt about disk and jet emission
from accreting black holes and neutron stars. A wide FOV monitoring towards the central Galactic bulge region  in the
SKA1 can demonstrate these
advantages in the early phase of SKA. 

\subsection{Monitoring of Extragalactic and Galactic X-ray transients}

Most of the high energy transients, such as GRBs and black hole transients,
are of the nature of rapid accretion onto compact objects in a large range
of mass accretion rate, from below $\rm 10^{-5}$ Eddington luminosity to
super-Eddington luminosity. X-ray monitoring observations in the past few
decades have usually reached a daily sensitivity of mCrab level, unable to
alert transient activities at earlier times when the X-ray flux is
substantially lower. At lower mass mass accretion rate regimes, the
observed electromagnetic flux is probably dominated by jet emission rather
than from the emission of the accretion flow itself, as suggested by the
observations of microquasars and the prediction of the accretion theory
such as advection-dominated accretion flow (ADAF) model for the accretion
flows in the lower mass accretion rate regimes (Esin et al. 1996). This
unambiguously highlights that the most sensitive and effective probe of
activities in black holes accreting at low mass accretion rate is through
the observations of their jet emission. Observations have shown that jet
emission in microquasars contribute to a broad wavelength from the infrared
(e.g., Russel et al. 2007) all the way up to at least ultra-violet (e.g.,
Yan \& Yu 2012; Degenaar et al. 2014). However, these jets are usually
detected in the radio band since only the most energetic ones could extend
to shorter wavelengths. This makes sensitive radio observations the most
promising probe of the activities of black holes accretion at low mass
accretion rates. 


\begin{figure}
\includegraphics[width=5.0cm,
height=10.0cm,angle=90]{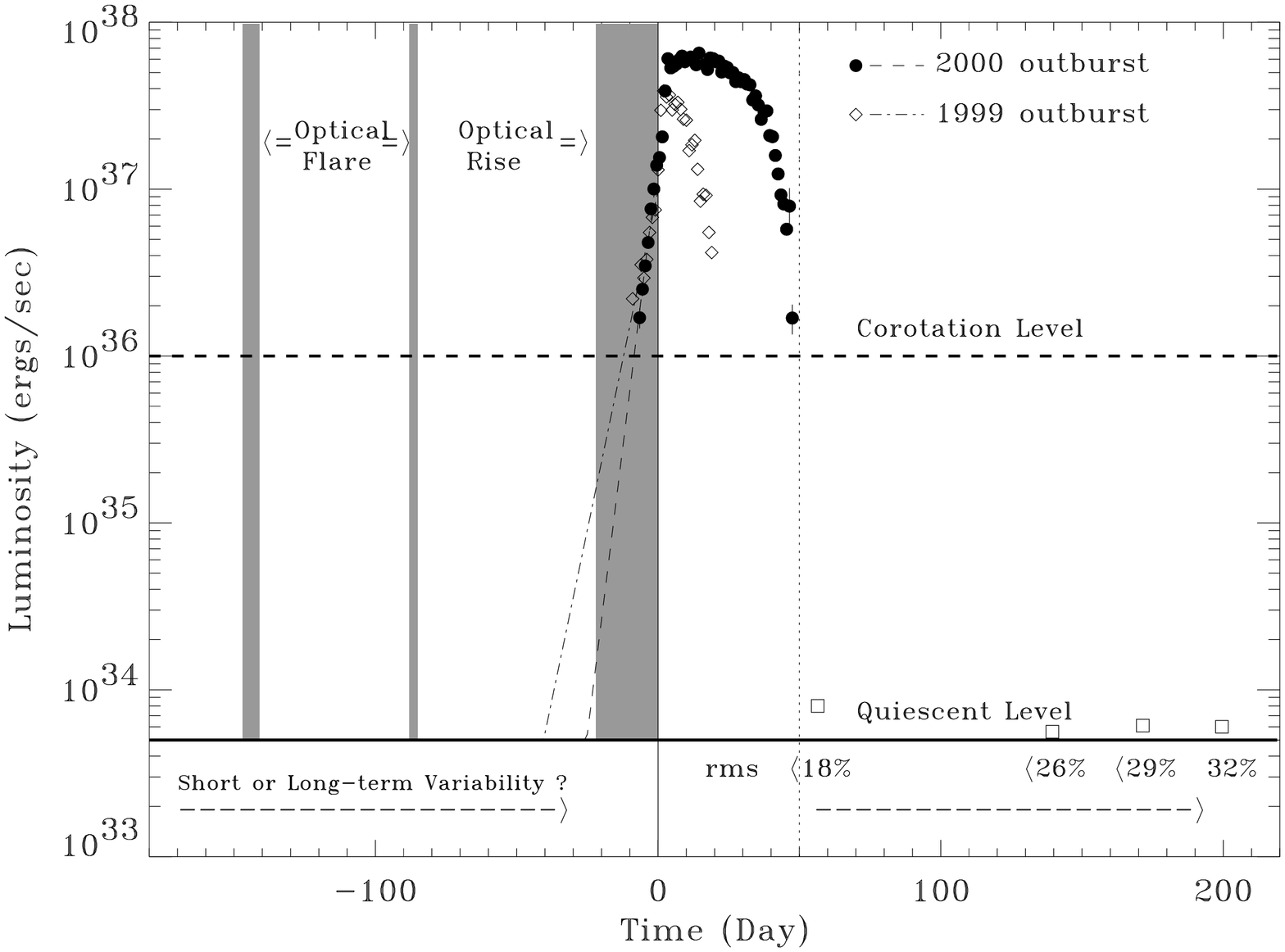}
\vspace{0.50cm}
\includegraphics[width=5.0cm, height=5.0cm]{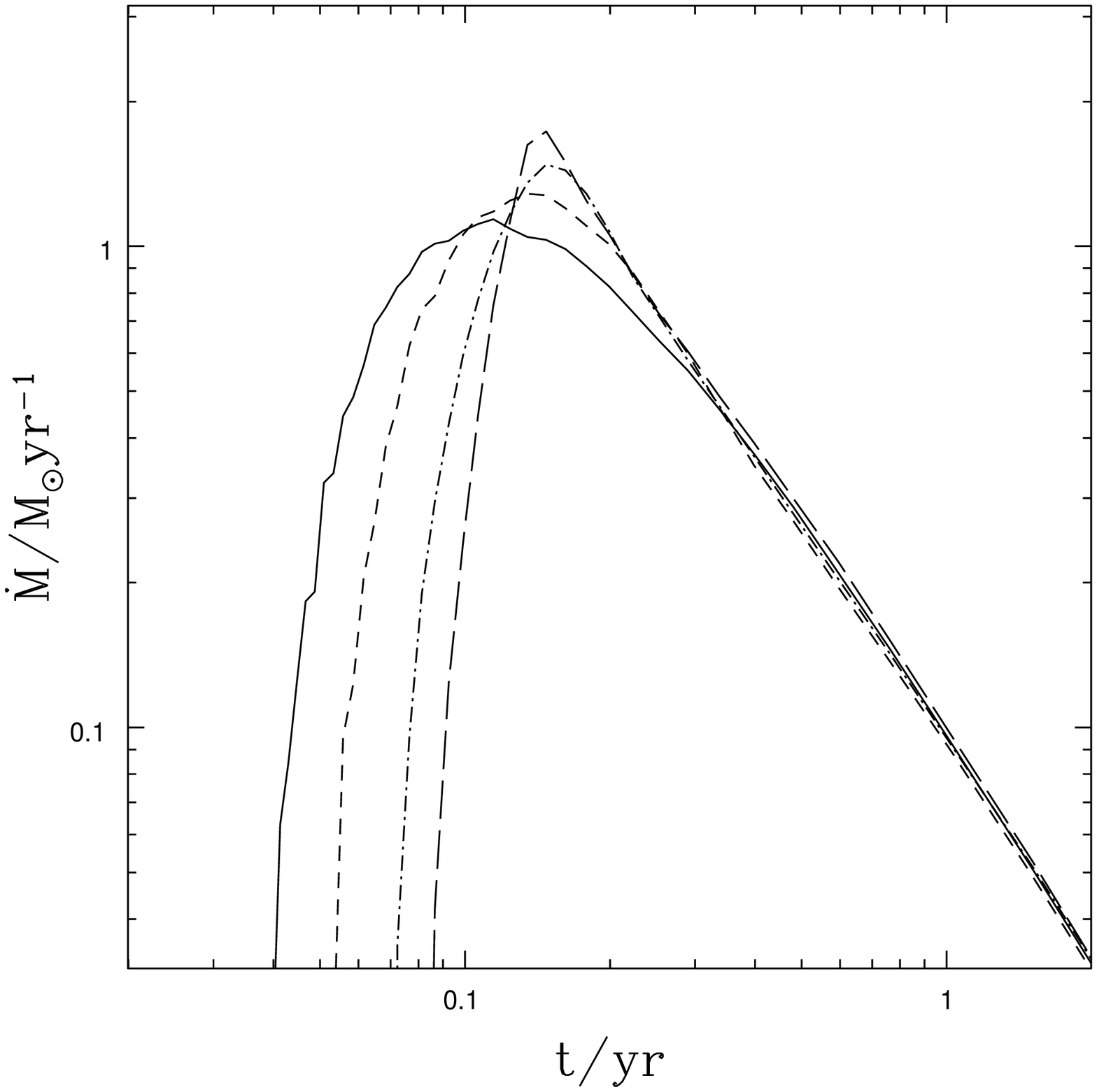}
\caption{Examples of X-ray outbursts from a Galactic X-ray binary transient
and the evolution of the fall-back rate expected from theory during typical
tidal disruption events (TDEs). {\bf Left:} An example of the X-ray
outbursts of the neutron star soft X-ray transient Aquila X-1 as observed
with the RXTE/ASM (2--12 keV). Typical rise time scale for black hole
transients are about twice longer. A linear extrapolation of the rising
trend in the X-ray flux back in time implies that outbursts started about
10--30 days from quiescent level before they were detected by the ASM on
board the RXTE. Future X-ray monitoring with Lobster-Eye instrument would
reach an X-ray sensitivity at the level of an X-ray luminosity of about
$5\times{10}^{34}$ ergs/s for this source, while SKA1 and SKA2 will be able
to monitoring this source immediately after the source rises from
quiescence, at the equivalent X-ray luminosity level of $1\times{10}^{34}$
ergs/s (corresponding to about 25 microJy in radio band at 8.5 GHz)
following the trend established from previous observations of a slope 0.6
(Miglari et al. 2011), and can detect such outbursts about a week before
any possible detections with the next generation all-sky monitors, leading
to a major step forward for the broadband monitoring of Galactic XRB
transients. {\bf Right:} The evolution of fall-back rate during typical
tidal disruption flares (from Lodato, King \& Pringle 2009). The typical
rise time is on the time scales of about 20 days (0.5 ${\rm T}_{m}$ as
calculated in Evans \& Kochanek 1989) and the rate-of-increase of the mass
accretion rate onto the supermassive black hole would be dramatic. Towards
lower fall-back rate back in time during the rising phase,
jet emission is expected to dominate when the corresponding mass accretion
rate is low (Yuan, Cui \& Narayan 2005). The SKA 2 has the sensitivity and
large-enough FOV to detect these events about a week before they reach the
maximum fall-back rates. Since typical peak luminosity of a TDE flare will
be super-Eddington for typical SMBHs of the order of $10^{6}$ solar masses,
the coverage of almost the entire rising phase with the SKA 2 will be
revolutionary. } 
\label{fig1}
\end{figure}

Besides observing transients in the radio wavelength, there are quite some
ambitious optical surveys with large FOV telescopes undergoing or
planned, which will bring significant progress on transient science. Among them, iPTF has carried out optical survey with
different cadences, from a 5-day cadence down to 90 seconds. Its followup
project ZWICKY will have a FOV of 47 square degrees with a survey speed of
3760 square degrees per hour. In its survey mode, Pan-SSTARS will cover
6000 square degrees per night, reaching a limiting magnitude of 24 in each
exposure of tens of seconds. On the other hand, LSST has a field-of-view of 9.6 square degrees
and will detect $\rm {10}^{6}$ transients per day. All these survey
telescopes will bring tremendous optical data on transients and persistent sources.
SKA's monitoring and survey capability provide opportunities for source
discoveries and synergetic multi-wavelength campaigns on newly discovered
transients and persistent sources. 

\section{New discovery window with the SKA: the early rising phase}

Space X-ray monitoring has been extremely successful in the past few
decades on high energy transients.
Some examples of the discoveries made with X-ray monitoring observations
include gamma-ray bursts and soft gamma repeaters, soft X-ray transient
outbursts, supernova explosions during shock breakout,  jetted tidal
disruption events, etc. Most often these all sky or large FOV X-ray
monitors are onboard satellites or space stations with a typical earth
orbit of 1.5 hours or so, and usually serves to send out alerts while
detailed X-ray spectral and short-term variability can not be simply
acquired. Detailed spectral and timing properties rely on follow-up
target-of-opportunity observations with dedicated space or ground
telescopes or experiments. In the past decades, space X-ray monitors have
reached a daily sensitivity at the mCrab level. In the next decade
the sensitivity of X-ray monitors would reach a flux level of about two
orders of magnitude lower, with the development of large field-of-view
Lobster-eye X-ray optics (Schmidt 1975; Angel 1979). 

When approaching black hole transients at lower luminosities, large FOV
radio facilities like SKA, in phase 2, will impose major challenge to the
role of X-ray all-sky of wide FOV monitoring, and will likely take the lead
instead according to our current knowledge of black hole accretion at low
luminosities. Gallo et al. (2003) showed that there is a correlation
between the radio flux and the X-ray flux in black hole X-ray binaries in
the hard state, with the power-law index of the radio flux being 0.7 of
that of the X-ray flux. The most updated data shows some obvious deviation
and large scatters from this correlation (e.g., Gallo et al. 2014).
However, black hole transients are still much more radio loud than neutron
star low mass X-ray binary transients (Migliari \& Fender 2006). It is
found that in the NS LMXBs the radio-X-ray correlation may follow
different pow-law slopes (Migliari et al. 2011). The transient NS LXMB
Aquila X-1 is typical for NS LMXBs showing a slope of 0.6, thus the slope
for Aquila X-1 type NS LMXBs can be used to give conservative estimate of
the LMXB transients as a whole. Together with the overall correlation in
black hole X-ray binaries, this implies that the radio flux decrease slower
than the X-ray flux when we approach lower luminosities. Therefore
opportunities with radio observations will start to arise; sensitive radio
observations will be able to surpass X-ray monitoring observations towards
lower mass accretion rates. Taking Aquia X-1 as the example (see Figure 1
left). The empirical correlation with a power-law index of 0.6 gives that
at an X-ray luminosity of ${\rm 10}^{34}$ ergs/s, which is just a little
above its quiescent luminosity, the corresponding radio flux at 8.5 GHz
would be 25 microJy. This radio flux level is well above the sensitivity of
SKA1-SUR or SKA1-LOW for 20 minutes observations or so while an X-ray
Lobster-Eye instrument can only reach the X-ray luminosity level of
$5\times{\rm 10}^{34}$ ergs/s with similar exposure time, which means the
next generation X-ray monitor is not able to detect it in the same period
of time. The same applies to accreting supermassive black holes. Merloni et
al. (2003) found that there is fundamental plane of black hole activity.
Similarly, the power-law index of the radio luminosity vs. the X-ray
luminosity relation is about 0.6. Again, this indicates towards lower
luminosities black hole accretors are relatively more radio bright compared
with their X-ray brightness, and therefore radio observations with the
SKA1-SUR or SKA1-LOW, and SKA2, will be more sensitive to source activities
at low luminosities (see Dewdney et al. 2013 for the SKA1 parameters). This
means SKA observations in radio is able to lead X-ray monitoring campaigns
on black hole transients of both Galactic and extragalactic origin.
Sensitive wide FOV monitoring in radio band with SKA hence opens a new
discovery window inaccessible with the X-ray monitoring of both stellar
mass black hole and supermassive black hole transients in the decades to
come. 

For stellar mass black hole transients in our Galaxy, the quiescent X-ray
luminosity in the 0.5--10 keV band is about $\rm {10}^{30-31}$ ergs/s
(e.g., Garcia et al. 2001). During the rising phase of black hole or
neutron star LXMB transient outbursts, the e-folding rise time scales in
the X-ray band measured with current X-ray monitors is about 3-5 days (Yan
\& Yan 2014, see early results of X-ray monitors in Chen et al. 1997). With
an equivalent X-ray sensitivity of 1-2 orders of magnitude better, the SKA1
(and the SKA2) design will be able to send out alerts of black hole
transient outbursts about a week ahead of X-ray monitoring detections if we
assume the rises of outbursts are of a simple monotonic
form. This can be seen in an example shown in Figure 1. Similarly,
extragalactic black hole transients such as tidal disruption events (i.e.,
solar-like star disrupted by previous dormant supermassive black holes
at the centers of normal galaxies) should also start to accrete matter from
the debris flow of a disrupted star from very low mass accretion rate to
super Eddington rate (Rees 1988; Evans \& Kochanek 1989; Lodato, King \&
Pringle 2009), although the exact accretion process in the early stages is
yet not well-studied. According to simple theoretical estimates, the
fall-back rate increases from quiescent level to super-Eddington fall-back rate up to more than 100
Eddington rate in about two-week's time during the rising phase of typical
TDE flares (Rees 1988; Evans \& Kochanek 1989; Lodato, King \& Pringle
2009). Taking a conservative estimate of the quiescent luminosity level of
supermassive black holes at centers of normal galaxies as $\rm {10}^{34}$
ergs/s (note: Chandra observations of the supermassive black hole source
Sgr A* at the center of our Galaxy gives such a value, see Baganoff et al.
2003; Nelsen et al. 2013), the mass accretion rate would rise by up to 10
orders of magnitude in about two weeks' time (Evans \& Kochanek 1989). The
e-folding rise time scale is then on the time scale of about ~ 0.5--2 days,
which can be ten time shorter than those seen during the rising phase of
outbursts from Galactic black hole LMXB transients. Therefore TDE flares
are cases of non-stationary black hole accretion in the extreme.

SKA has a very large field of view. For example, at 1 GHz, its
field-of-view of SKA2 is around 200 square degrees. Both SKA1 and SKA2 are
also very sensitive. For instance, SKA1-SUR's 20 minute sensitivity at 1
GHz is at 0.01 mJy level, and SKA2's 20 minute sensitivity at 1 GHz is at
the $\mu\rm Jy$ level. Due to its high sensitivity, its convenience for
observations of our Galaxy and its large FOV, SKA is the best astronomical
facility to catch the rising phase of outbursts or flares from accreting
stellar mass black holes and supermassive black holes. Little is known
about the early rising phase of these outbursts or flares from previous
observations due to insufficient sensitivity of current and past space
X-ray monitors. SKA1-SUR and the SKA Phase 2 will open a huge discovery
window in catching the very early rising phase of those black hole
transient flares and outbursts, including both black hole binary
transients and tidal disruption flares. Our estimates based on current
knowledge of soft X-ray transient outbursts and TDEs suggest that SKA1,
through a dedicated Galactic bulge program, is able to detect black hole
binary transients in our Galaxy; while SKA2 is able to detect extragalactic
tidal disruption events during the early rising phase about a week ahead of
the most sensitive X-ray monitors in the next decades (see Figure 1 left and right).

\section{Accretion physics and jet activity in extreme non-stationary accretion regimes}

Observations of X-ray spectral states in Galactic X-ray
binaries have shown that the primary parameter - the mass accretion rate is
not the only parameter which drives spectral state transitions (Miyamoto et
al. 1995; Homan et al. 2001; Maccaroni \& Coppi 2003). The hard-to-soft
spectral state transition usually observed during the rising phase of
outbursts actually occurs at  luminosities which correlates with the peak
flux of an outburst or flare (Yu et al. 2004; Yu et al. 2007; Yu \& Dolence
2007), due to the rate-of-increase in the mass accretion rate, which is not
negligible,  and in most of the spectral state transitions seen in bright
Galactic X-ray binaries the effect of the rate-of-change of the mass
accretion rate on the transition luminosity dominates (Yu \& Yan 2009;
Tang, Yu \& Yan 2011). These studies indicate that the accretion properties
are set up early, and one can predict spectral state transition ahead of
time by measuring the rate-of-increase of the X-ray luminosity (Yu \& Yan
2009). It is worth noting that our current knowledge of accretion regimes
are actually based on stationary accretion theories. Non-stationary accretion 
regimes are largely not explored. The studies of the hysteresis effect
implies that there is actually a non-stationary accretion regime, e.g., the black hole hard state
regime corresponding to non-stationary accretion which exists above the
mass accretion threshold below which the classical hard state sits as
explained by existing theory (e.g., Esin et al. 1997). Because there is
certain complex correlation between the radio flux and the X-ray flux in
the hard spectral state (Gallo et al. 2003; but with so-called outliers),
there is also evidence that non-stationary accretion plays an important role in
powering the radio jet. This is suggested by a possible detection of
the correlation between the rate-of-change of the mass accretion rate and
the episodic jet power (Zhang \& Yu, 2014, see Figure 2).  Since both TDE
flares and black hole binary transients can be detected by the SKA at much
earlier times and at much lower luminosities, SKA will provide the
opportunity to probe jet activity in the non-stationary accretion regimes
over a radio flux range by more than 4 orders of magnitude, which roughly
corresponds to an X-ray flux range by about 7 orders of magnitude in
the Galactic black hole transients. Stronger effect of non-stationary accretion
is expected in extragalactic X-ray transients associated with tidal
disruption events due to their much shorter e-folding rise time scale. Our
conservative estimate suggest that TDE flares can rise in luminosity as large
as 10 orders of magnitude on the time scale of 20 days, accompanied by an
extremely large rate-of-change of the mass accretion rate inaccessible in
other black hole systems.

\begin{figure}
\includegraphics[width=8.0cm,height=5.0cm]{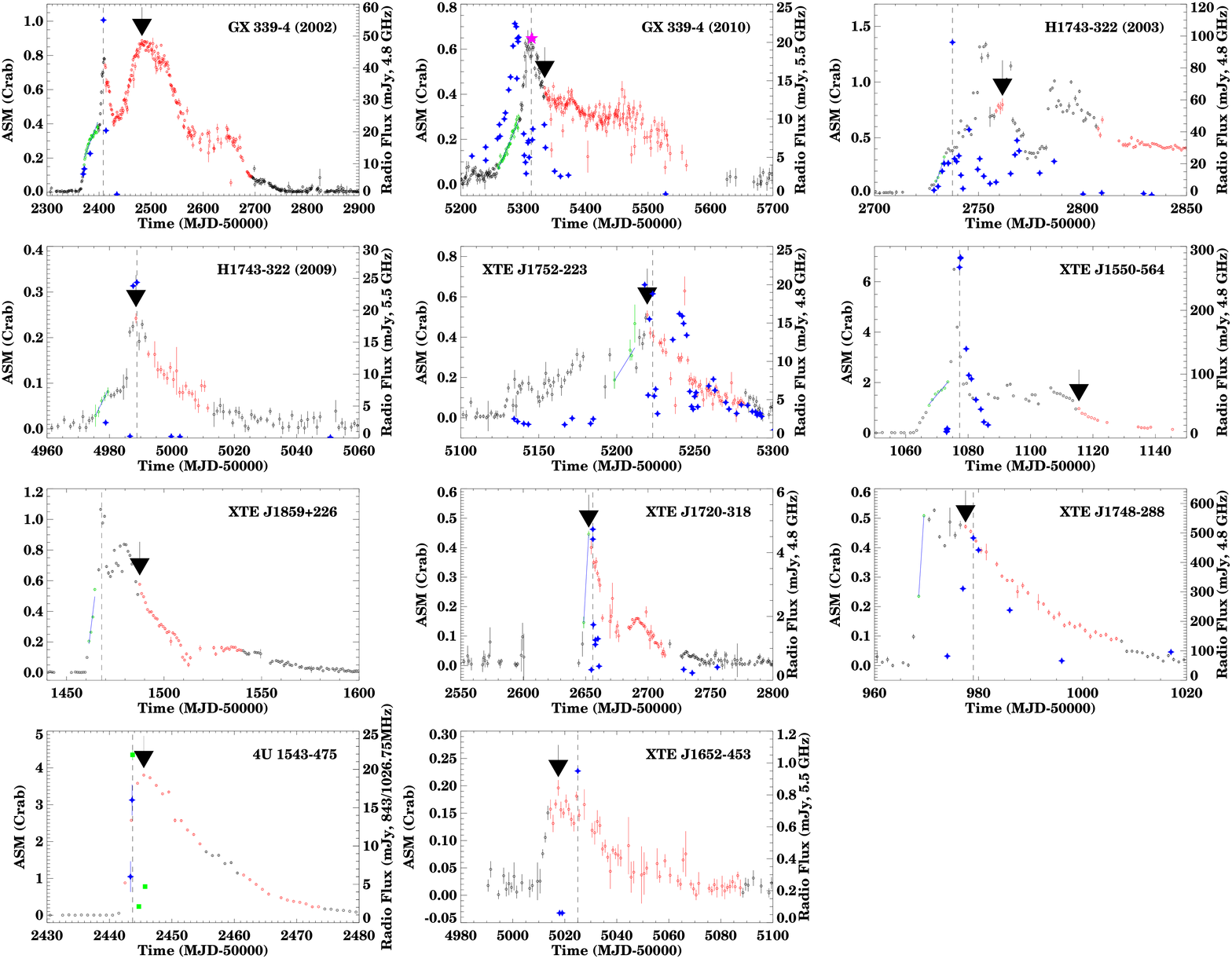}
\includegraphics[width=7.0cm,height=5.0cm]{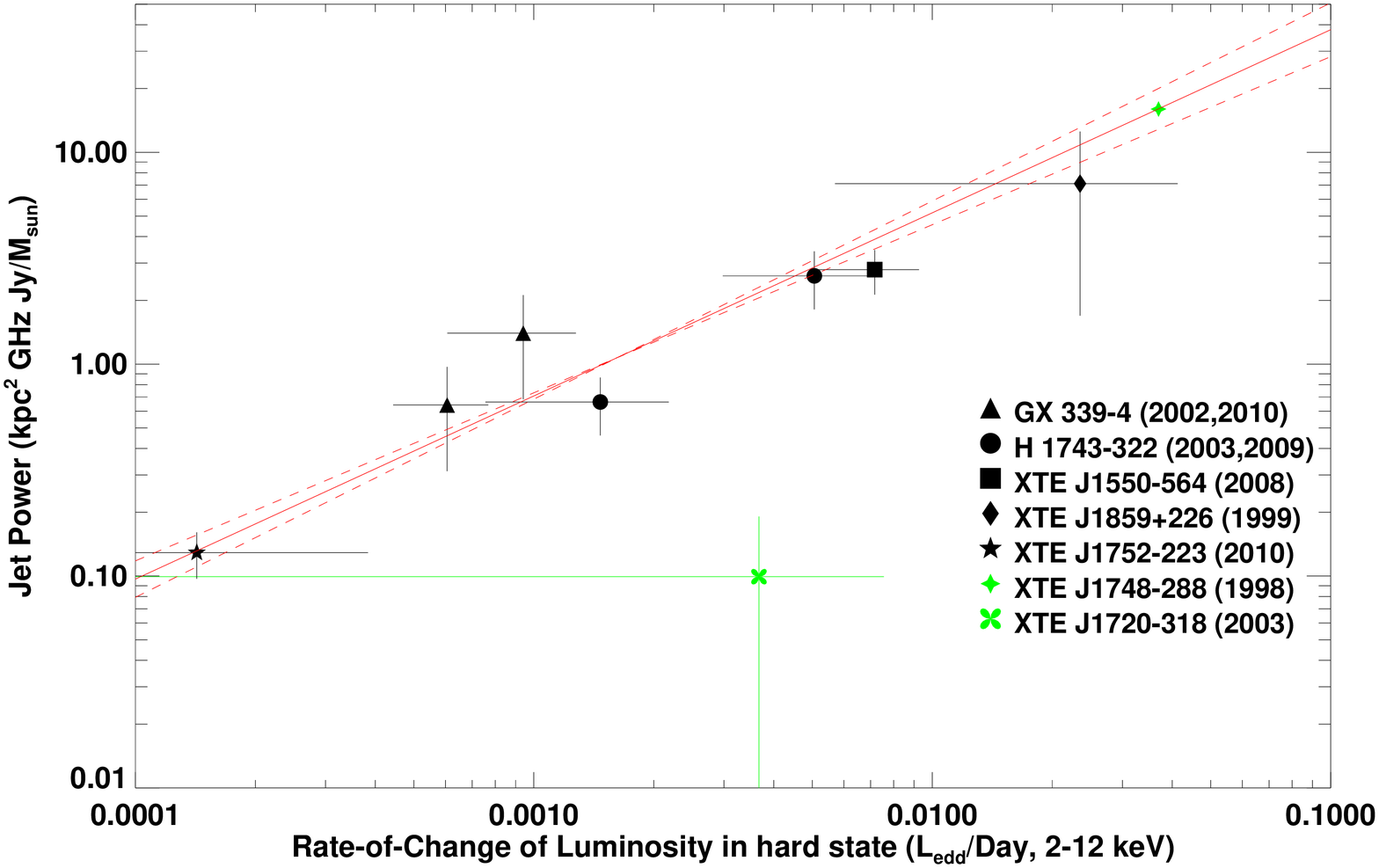}
\caption{Evidence of the role of non-stationary accretion on the power of
episodic jets in microquasars (figures adapted from Zhang \& Yu 2014). {\bf
Left:} X-ray outbursts of black hole transients with radio observations of
the episodic jets. Data shown in blue are radio measurements, data shown in
red correspond to black hole soft state. Data marked in green correspond to
the segments from which the rate-of-increase of the X-ray luminosity is
measured for the rising phase of the corresponding outburst. {\bf Right:}
The relation between the episodic jet power and the rate-of-change of the
X-ray luminosity measured before the hard-to-soft state transition during
black hole transient outbursts (see Figure on the left). The dark points
corresponding to sources with good measurements or estimates of mass,
distance, X-ray and radio fluxes, while the green points indicate sources
with large uncertainties.The best curve is over-plotted in red, with the
ranges of uncertainty shown as  dotted lines.} 
\label{fig2}
\end{figure}

\section{Recommended campaigns on extragalactic and Galactic black hole
X-ray transients}

As we described above, wide FOV sensitive observations in the radio wavelength
have great advantages in detecting stellar mass and supermassive black hole
X-ray transients during the very early rising phase of their outbursts or
flares. Using SKA1-SUR band 2 with a FOV of 18 and a sensitivity of
0.351mJy/s$^{-1/2}$, in order to reach 0.01 mJy/beam per field, we would
need 20.5 minutes observation. The predicted SKA2 sensitivity is about 10
times sensitive than SKA1, which means we can reach the same sensitivity
per field in only 12 seconds. With an increased FOV of SKA2, an all-sky
survey down to 0.01 mJy/beam per field is applicable. 

In Phase 1, SKA1-SUR has a FOV of about 18 square degrees in PAF band 2.
For Galactic black hole or neutron star transients, in order to cover the
very early phase of any future outbursts of Galactic black holes as well as
neutron star LMXB transients, most of which reside in the Galactic bulge, a
dedicated monitoring campaign of the Galactic bulge is recommended on the
daily basis with the SKA1-SUR at 1 GHz. Past X-ray observations of the
Galactic bulge indicate that there are about 80 known persistent and
transient X-ray binaries in the central 25 x 25 degrees of the Galactic
bulge, and probably hundreds to discover in the future. In a smaller
central Bulge region of 8 x 8 squared degrees, there are 35 known
persistent and transient X-ray binaries. In order to cover this region, we
recommend that SKA1-SUR performs more than 4 short observations of the 8 x
8 square degrees of the Bulge down to 0.01 mJy flux level. According to the
empirical relation established from radio flux vs. X-ray flux correlation,
these SKA1-SUR observations will be able to detect outbursts of Galactic
black hole or neutron star LMXB transients before they are visible to
current or future X-ray monitors, and will be especially successful in
detecting accretion-powered millisecond pulsars which is usually
radio-louder than the Atoll type neutron star LMXBs. After the
implementation of the full version of the SKA, because the FOV would reach
200 square degrees at 1 GHz, three observations will be able to cover the
central Bulge region. 

With full SKA's capability of a FOV of about 200 square degrees, SKA is a
powerful machine to detect new tidal disruption events at very early times.
This will bring tremendous opportunities for Target-of-Opportunities (ToO)
with space X-ray observatories such as Athena+ and ground optical
telescopes on the scale of tens of meters, since there will be plenty time
for both ground and space observatories to respond (about 10 days' earlier
than current X-ray monitoring can provide), which will push time domain
astrophysics on black hole transients, especially the tidal disruption
events, to the extreme regimes. However, there are difficulties for the
SKA1-SUR or the SKA1-LOW to sufficiently detect some extra-Galactic
transients such as tidal disruption events due to their limited FOV, but
still there are occasions when tidal disruption events occur in the
SKA1-SUR or the SKA1-LOW's FOV. A better strategy during SKA1 for black
hole transient science is to use SKA1-SUR and SKA1-LOW to perform
quasi-simultaneous observations of the same field of view of optical
surveys or X-ray all-sky monitoring, such as the LSST's FOV as much as
possible. In this way, due to broadband measurements, empirical
classification of transient events could be made straight forward. This
would allow SKA-MID to perform sensitive follow-up observations of some
candidate tidal disruption events timely. 

In summary, the motivations of SKA1 and SKA2 campaigns on Galactic and
extragalactic black hole X-ray transients are 1) the coverage of the very
early rising phase of outbursts or flares and sending out extremely early
alerts; and 2) rather complete coverage of jet activities in the 
non-stationary accretion regimes corresponding to a large range of the mass
accretion rate and its rate-of-change, which are inaccessible in other
accreting black hole systems. 

\section{Scientific outcomes of SKA1 and SKA2}

We expect the following outcomes of the SKA1 and SKA2 on the science of
black hole transients containing both stellar mass or supermassive black
holes. 

In the SKA1, a Galactic bulge monitoring campaign with SKA1-SUR in the
central 8 x 8 squared degrees for 1.5 hours on the daily basis will be able
to cover the very early rising phase of nearly half of the stellar mass
black hole and neutron star LMXB transient outbursts, which will lead the
detection in the X-rays by sensitive X-ray monitoring observations. Such a
Galactic bulge campaign will also detect some APMSP outbursts. Such a
monitoring campaign will answer how jet is powered in the non-stationary
regimes in a large range of mass accretion rate at the same time of the
significant rate-of-change in the mass accretion rate, which will address
accretion and jet physics in the non-stationary accretion regimes in LMXB
transients. The program will also make SKA1-SUR likely the first facility
to detect new black hole and neutron star LMXB transients in the Galactic
Bulge at unprecedented low flux level and will be able to send out alerts
on new BH or NS transients or outbursts to other ground or space
observatories. On the other hand, only coordinated or follow-up SKA1
observations of extragalactic TDE flares are possible in SKA1. 

In the SKA2, an efficient monitoring of the Galactic bulge of a 25 x 25 FOV
can be achieved at 1 GHz. Additionally, SKA2 will provide the best
opportunity to study extragalactic SMBH transients such as TDEs. The most
significant scientific impact on the study of black hole transients would
come from the detection of a lot extragalactic TDE flares in the early
rising phase with SKA. The SKA2's sensitivity allows detections of such
events at the very early times, which is mostly unknown in both theory and
observation. SKA2 would provide a complete coverage of their jet activities
over an extremely large range of mass accretion rate and its
rate-of-change, and is able to send out alerts extremely early for the
largest ground and space observatories to respond. Thus SKA will provide
the golden opportunity to solve important astrophysical questions
associated with black holes transients in binaries and at the centers of
normal galaxies.


\begin{thebibliography}{*}

\bibitem{} Angel, J. R. P. 1979, ApJ, 233, 364

\bibitem{} Bloom, J.~S., Giannios, 
D., Metzger, B.~D., et al.\ 2011, Science, 333, 203

\bibitem{2011Natur.476..421B} Burrows, D.~N., Kennea, 
J.~A., Ghisellini, G., et al.\ 2011, \nat, 476, 421

\bibitem{1997ApJ...491..312C} Chen, W., Shrader, C.~R., 
\& Livio, M.\ 1997, \apj, 491, 312

\bibitem{2013MNRAS.428.2500C} Corbel, S., Coriat, M.,
Brocksopp, C., et al.\ 2013, \mnras, 428, 2500

\bibitem{} Dewdney, P., Turner, W., Millenaar, R., McCool, R., Lazio, J.,
Cornwell, T., 2013, ``SKA1 System Baseline Design'', Document number
SKA-TEL-SKO-DD-001 Revision 1

\bibitem{1989ApJ...346L..13E} Evans, C.~R., \& Kochanek, C.~S.\ 1989, \apj, 346, L13

\bibitem{2004NewAR..48.1399F} Fender, R.\ 2004, \nar, 48, 
1399

\bibitem{2003MNRAS.344...60G} Gallo, E., Fender, R.~P.,
\&amp; Pooley, G.~G.\ 2003, \mnras, 344, 60

\bibitem{2014MNRAS.445..290G} Gallo, E., Miller-Jones, 
J.~C.~A., Russell, D.~M., Jonker, P.~G., Homan, J., Plotkin, R.~M., 
Markoff, S., Miller, B.~P., Corbel, S., Fender, R.~P.\ 2014.\mnras, 445, 
290. 

\bibitem{} Garcia, M. R.; Murray, S. S.; McClintock, J. E.; Narayan, R. 2001, ApJ, 553, L47

\bibitem{2009MNRAS.398.1392L} Lodato, G., Nayakshin, 
S., King, A.~R., \& Pringle, J.~E.\ 2009, \mnras, 398, 1392 

\bibitem{2005A} Merloni, A., Heinz, S.; Di Matteo, T.\ 2005, \apss, 300, 45


\bibitem{2003MNRAS.345.1057M} Merloni, A., Heinz, S.; di Matteo, T.\ 2003, \mnras, 345, 1057

\bibitem{2006MNRAS.366...79M} Migliari, S., Fender, R.~P.\ 2006.\mnras, 366, 79.

\bibitem{2011MNRAS.415.2407M} Migliari, S., Miller-Jones, J.~C.~A., Russell, D.~M.\ 2011.\mnras, 415, 2407. 

\bibitem{1988Natur.333..523R} Rees, M.~J.\ 1988, \nat, 333, 523 

\bibitem{2007MNRAS.379.1401R} Russell, D.~M., Maccarone, T.~J., K{\"o}rding, E.~G., Homan, J.\ 2007.\mnras, 379, 1401

\bibitem{2008Sci...321..223S} Schawinski, K., 
Justham, S., Wolf, C., et al.\ 2008, Science, 321, 223

\bibitem{} Schmidt, W. K. H. 1975, Nucl. Inst. Methods, 127, 285

\bibitem{2011RAA....11..434T} Tang, J., Yu, W.-F., 
\& Yan, Z.\ 2011, Research in Astronomy and Astrophysics, 11, 434 

\bibitem{} Yan, Z. \& Yu, W. 2014, submitted to ApJ (aarXiv: 1408.5146)

\bibitem{2007ApJ...667.1043Y} Yu, W., \& Dolence, J.\ 2007, \apj, 667, 1043 


\bibitem{2007ApJ...663.1309Y} Yu, W., Lamb, F.~K., Fender, 
R., \& van der Klis, M.\ 2007, \apj, 663, 1309 


\bibitem{2004ApJ...611L.121Y} Yu, W., van der Klis, M., 
\& Fender, R.\ 2004, \apj, 611, L121 

\bibitem{2009ApJ...701.1940Y} Yu, W., \& Yan, Z.\ 2009, \apj, 701, 1940 

\bibitem{2005ApJ...620..905Y} Yuan, F., Cui, W., Narayan, R.\ 2005.\  \apj, 620, 905. 

\bibitem{} Zhang, H. \& Yu, W., 2014, submitted to MNRAS

\end{thebibliography}
\end{document}